# Cuba: the strategic choice of advanced scientific development, 1959-2014

Angelo Baracca* and Rosella Franconi°

*Department of Physics and Astronomy, University of Florence, Italy, baracca@fi.infn.it
°ENEA, Italian National Agency for New Technologies, Energy and the Environment, Department for Sustainability, Casaccia Research Centre, Rome, Italy, rosella.franconi@enea.it

**Abstract**

Cuba is continuing attracting the attention of the international scientific community for some important and unexpected achievements in applied science such as health biotechnology. They represent outcomes of the 1959 decision of Cuba to develop an advanced scientific system in order to address the most urgent problems for the development of the country and to overcome the condition of subalternity. This ambitious objective was tackled in a very original way, making a broad and wide-ranging recourse to every effective support and collaboration, with Soviet but also Western scientists and institutions, in addition to a peculiar Cuban inventiveness. Indeed, immediately after the revolution, Cuba developed an advanced and articulated scientific system, and achieved a level of excellence in leading scientific fields, like biotechnology, quite independently from the Soviet Union, which was behind in this field.

Even the collapse of the Soviet Union in the early 1990s, that could have put the achievements of the Revolution at risk, posing again the threat of subalternity, under an intentionally worsened American embargo, did not change this trend: once more Cuba addressed this challenge reconfirming the strategic choice of supporting its most advanced and profitable scientific sectors, especially the capital-intensive and typically American field of health biotechnologies. This strategy proved to be once again a well-chosen course of action.

## An original approach: a twofold successful fight against subalternity

Overcoming the situation of subalternity is the scary challenge that every developing country has to address. Defining what a situation of autonomy really means is also a complex problem. Sometimes a developing country achieves such a condition inside the sphere of influence of a leading power, passing in some sense from a complete lack of autonomy to a dependence from the latter, and its fortunes. Not infrequently overcoming subalternity is a result far from being a definitive achievement, as it may be illustrated by Spain after the Spanish-American war of 1898, or Russia at the unexpected collapse of the Soviet Union

Actually, in its recent history, after the victory of the Revolution in 1959, the Caribbean island met the challenge of overcoming subalternity in two occasions, and in both cases, in very critical conditions and a strongly disadvantageous international situation, addressed it in very original forms. In both cases the basis of Cuba's strategy was to rely on the development, in the second case a relaunch, of an advanced scientific system.

*The first challenge for Cuba, overcoming subalternity through scientific development*

In 1959 the perspectives for a modern scientific system appeared as very remote and fanciful. Cuba was a prevalently rural country, with scarce natural resources that were strongly controlled by foreign interests, and a population of barely 7 million inhabitants, with an official literacy rate between 60-76%, largely because of lack of educational access in rural areas and a paucity of instructors (Kellner 1989, 61). Actually, the country had a fairly good system of secondary instruction, and of basic scientific education, but even in the university did not include modern scientific disciplines, such as quantum theory or genetics. As the 2010 UNESCO Science Report recalls,

> «A report by the *ad hoc* Truslow Commission of the International Bank for Reconstruction and Development, which had traveled to Cuba to study the provision of loans, stated unequivocally in 1950 that 'in the field of applied research and labs, there was no development at all in Cuba' (Sáenz and García-Capote 1989)» (Clark Arxer 2010).

Aiming unwaveringly for the development of an advanced scientific system, for a small country in these backward conditions, could have appeared as a definitely unrealistic objective. On the contrary, this backwardness was overcome in such a surprisingly short period of a couple of decades. At the turn of the 1980s Cuba rivaled the percentage of university graduates and physicians, and the overall level of their scientific training in many highly developed nations, and had no equals among other underdeveloped countries (Hoffmann 2004, 166-168).

Even more peculiar was how Cuba faced this endeavor, in very original forms, essentially with an exceptionally open-minded attitude, resorting to supports and advice from every side (from both Western and Soviet scientists and institutions, depending on their suitability and usefulness), and putting in the first place in its scientific choices the social and economic priorities of the population and the country. Emblematic examples are how Cuba addressed disaster management and mitigation, and its health situation[1]. In spite of the expatriation of almost one half of doctors after the Revolution, Cuba addressed and defeated the Third World illnesses during the 1970s, and soon reaching a First World health profile. On this basis biotechnology had an impetuous take-off, centered on domestic problems, soon reaching by the turn of the 1980s an international standard, all the more impressive for this small and resourceless island in such a capital-intensive sector which is the quintessence of American industry and supremacy (Baracca and Franconi, 2016). Such an astonishing result is acknowledged by the most influential scientific journals, like *Nature, Science*, and the specialized magazines (Kaiser 1998; Thorsteinsdóttir et al. 2004; Buckley 2006; López Mola et al. 2007; Evenson 2007; Editorial 2009; Starr 2012; Fink et al. 2014).

> «Cuba's outstanding achievements in health biotechnology are a source of inspiration for the developing world. They are all the more impressive considering that the island is a small, relatively poor country that has suffered serious economic difficulties for more than a decade. ... Despite these difficulties, Cuba's strong and continued emphasis on science since shortly after the 1959 revolution has resulted in a highly developed health biotechnology sector» (Thorsteinsdóttir et al. 2004, 19).

It is true that the country enjoyed the substantial support from the Soviet Union and the Socialist countries. However, this is far from fully explaining Cuba's success in crucial advanced scientific sectors. In physics, the Soviet support had undoubtedly a role that could hardly be underestimated (Baracca, Fajer and Rodríguez 2014): Soviet physics had an outstanding international standard, and the great majority of the Cuban physicists has studied or specialized in the Soviet Union or the Socialist countries, where they had access to the most elitist scientific institutions. Nevertheless, as we shall see, since the early 1960s a considerable number of physicists from Western countries developed collaborations with Cuba, and has brought instrumental contributions in the advances in some crucial fields, such as microelectronics; moreover, the number of Cuban physicists who were educated in Italy, France or elsewhere was certainly smaller but fairly significant. But the situation was radically different in the biological sciences. For complex historical reasons Soviet biology did not keep up with the most innovative developments in genetics and molecular biology. It was however precisely in these fields that Cuba reached the most brilliant achievements starting from the 1980s, and these results were completely independent from Soviet science (Reid-Henry 2010; Baracca and Franconi 2016, Chapter 5). In fact, as we shall discuss below, the training of a whole generation of Cuban geneticists and molecular biologists entirely depended on intensive courses carried out by Western, mainly Italian, biologists in Cuba during the early 1970s (Baracca and Franconi 2016, pp. 47-50), and the successive stages and collaborations of the Cuban fellows in some of the most advanced international laboratories.

All these contacts and collaborations were rapidly integrated in the Cuban growing scientific system, which reached its maturity and its most brilliant achievements during the 1980s. Cuba had overcome its situation of subalternity, acquiring a substantial scientific autonomy, which does not mean isolation, let alone autarky, which is a nonsense in modern science, but the ability of interacting and collaborating on an equal footing with the most rated international colleagues and laboratories.

---

[1] A further recent striking example of this choise: Cuba has become the first country to earn certification by the World Health Organization (WHO) and United Nations officials to achieve elimination of mother-to-child transmission of HIV and syphilis. http://who.int/mediacentre/news/releases/2015/mtct-hiv-cuba/en/#

*The spectrum of subalternity arises again*

The Cuban scientific system concretely proved the strength and compactness it had reached when it held out against the almost sudden and unexpected disappearance of the Soviet Union. Nevertheless, the shock was dramatic, for the whole Cuban economy and society. Not only the country lost from one day to the next a basic economic support,[2] but it had to abruptly pass from a protected market to one submitted to the capitalistic rules, such as the Agreement on Trade-Related Aspects of Intellectual Property Rights (TRIPS). As we have remarked, subalternity is not overcome once and for all. Could Cuba resist and overcome this dramatic situation? Many observers predicted the forthcoming downfall of the Cuban regime, as it happened for all the remaining East European ones. After three decades Cuba faced again the challenge of meeting a renewed danger of subalternity, let alone survival, in a radically changed world. But Cuba seems to defy all odds. Once again the country addressed the challenge reconfirming the same recipe, that is reinforcing the support to selected scientific sectors, primarily its biotechnology complex, as a strategic choice in order to sustain its economy and its social achievements. Other scientific sectors, which had also reached a high international standard, were seriously jeopardized by the critical economic situation, but were not abandoned, and, while biotechnology has made further brilliant achievement, the Cuban scientific system in its whole withstood the blow.

As the already cited 2010 UNESCO Science Report states,

> «By the dawn of the 21st century, Cuba was perceived as being a proficient country in terms of scientific capacity, despite having experienced more than four decades of a trade embargo and restrictions on scientific exchanges imposed by successive US administrations … In Latin America and the Caribbean, only Brazil and Cuba qualified as 'proficient'» (Clark Arxer 2010).

In this paper we will briefly examine this challenging case, of how Cuba has overcome the initial condition of subalternity choosing as a key objective an advanced scientific development, and which were the main features of this process. Our reconstruction stops at 2014 since the re-establishment of the relations with the United States will presumably introduce unpredictable changes.

**A future of men and women of science for the new revolutionary society**

The revolutionary leadership faced enormous problems of every kind, while the situation with the near imperial power was getting out of control and the relation with the USSR was not yet working out. All the more a rhetoric gamble could appear Fidel Castro's famous bold statement in January 1960, just one year after the victory of the Revolution:

> «The future of our country has to be necessarily a future of men [and women] of science, of men [and women] of thought because that is precisely what we are mostly sowing; what we are sowing are opportunities for intelligence» (Castro 1960).

Who could have said that this was precisely what happened in Cuba within barely 2 or 3 decades? Although probably in a form that not even Fidel would have imagined: on the other hand, who could foresee the future developments of science?

> «After the 1959 revolution, Cuba made it a priority to find new ways to care for a poor population; part of the solution was training doctors and researchers» (Starr 2012).

---

2 Indeed, the Cuban-Soviet relations had gradually deteriorated in the second half of the 1980s, particularly under Gorbachev, both for ideological reasons, and since the latter strongly tightened its purse strings in the last years. On the other side, Cuba never completely got along with the Soviet system, as we have already remarked for its scientific choices. Nevertheless, reliance on although Soviet support and the COMECON market had always been a sound basis for Cuban economy.

The very young but learned revolutionary leadership[3] had a clear view not only of the crucial role of culture and science for the emancipation of the people and the liberation from subalternity,[4] but also of which fields owed to be developed to this end. Public healthcare became immediately a priority of the revolutionary government. 'Che' Guevara supported a strong development of basic science, as Minister of Industry promoted courses of mathematics for the employees of the Ministry, and foresaw the role of electronics and automation as leading fields and "fundamental political tasks for our country" (Pérez Rojas 2014, 282; Baracca, Fajer and Rodríguez 2014, 149). In 1961 he was the main promoter of sending Cuban students to graduate in the Soviet Union, the first six to study electronic engineering, then moved with his assent to follow a physics degree course (Baracca, Fajer and Rodríguez 2014, 137).

The post-revolutionary environment was pervaded by great enthusiasm and deep ferments and forms of participation, creativity and inventiveness. After the 1961 widespread literacy campaign, which almost eradicated illiteracy, the government laid the foundations for a profound reform of higher education, with free enrollment for all eligible students (Arias De Fuente 2014). At a time of violent aggression from abroad, the 1962 Higher Education Reform Law was a crucial step, which arose from a lively discussion that involved university professors, outside professionals and students, and envisaged a modern university (still conceived on western models) and a modern scientific system in which teaching was closely related to scientific research. Every scientific development since then was planned with the explicit aim of developing research programs, having as its priorities the basic social and economic needs of the country.

*A priority to physics*

Concerning scientific development, a peculiar choice characterized the process in Cuba, inasmuch the development of a strong and multi-purpose physics sector was considered as a strategic priority, a sound basis for all other scientific and technical fields. This choice proved correct, since physics has acted as the backbone of intellectual life and has provided many other fields with scholars, methods and scientific approaches that have been important for the further development of other disciplines in Cuba, such as medicine, biotechnology and the nanosciences.

The Faculty of Science, which previously included three sections (*Ciencias Físico-Matemáticas*, *Físico-Químicas*, and *Naturales*) whose task was the preparation of high school teachers, in 1962 was articulated into seven Schools, of Mathematics, Physics, Chemistry, Biological Sciences, Geology, Geography, and Psychology. In each of them 5-year professional degree courses were created. The situation in the new Schools was extremely difficult, since they initially lacked both material resources and professional staff (many teachers left the country after the Revolution).

The School of Physics coped with these problems in highly original ways, combining native resourcefulness with the pursuit of all possible kinds of support and collaboration. The students were actively involved in teaching activities, those in the final years of their studies being used as "assistant students" to teach freshmen (Baracca, Fajer and Rodríguez 2014, p. 59). "Western" textbooks considered most adequate for the purpose were reproduced and made freely available to teachers and students as so-called "Revolutionary Editions", thus circumventing the economic embargo enforced by the US government.

Moreover, the Cuban Revolution had raised a lively interest and great hopes all around the world for its original features. Cuba was destination of visits of professionals and intellectuals from all around the world, wishing to bring their advice, concrete support and collaboration. Among the most famous one can mention Jean-Paul Sartre (Sartre 1961), Simone de Beauvoir, Leo Huberman and Paul Sweezy (Huberman and Sweezy 1960), Charles Wright Mills (1916–1962),[5] the French Marxist economist Charles Bettelheim, and Michal Kalecki from Poland. The Cuban society and leadership were very receptive to foreign advice at those times, and Cuba became a crucible of lively contributions and experiences.

---

[3] In 1959 Fidel Castro was 33, Ernesto 'Che' Guevara 31, Raul Castro 28, Camilo Cienfuegos 27.

[4] José Martí (1853-1895) had been the first Latin American who clearly developed a full consciousness of the indissoluble tie between political independence, education and emancipation from subalternity, and had specifically proposed science education, the study of nature, as an instrument for individual autonomy and the promotion of social progress, because "to study the forces of nature and learn to control them is the most direct way of solving social problems" (J. Martí J. Obras completas. Havana, Edición del Centenario, Editorial Lex, 1953, I, 1076).

Almost contemporaneously to the arrival of the first Soviet specialists, quite a few Western physicists visited the School of Physics since 1961 for more or less extended periods of time, even year, and brought a vital support to give advanced courses, organize laboratories and workshops, and foster the early research activities.

Around mid-1960 a lively debate about the choice of the fields of physics to prioritize along with the most urgent problems of the country, involved Cuban and foreign scientists, who participated to the 1968 Cultural Congress of Havana. While the favourite "forefront" fields were worldwide nuclear and high-energy physics[6], the choice in Cuba finally fell on solid-state physics, a field in which the United States had the leadership and were developing an advanced industrial sector. Solid-state and semiconductor physics was given priority at the School of Physics of the University of Havana (in 1967 the first solid-state diode was made in Cuba in collaboration with Western visiting professors) but the other important fields were not disregarded, and other scientific centres were created.

In 1970 a peak of graduations in physics was reached, which provided a critical mass of well trained physicists that could feed the needs of the growing number of centres and scientific disciplines.

*Growth of the Cuban Scientific System*

In 1962 the Cuban Academy of Science was revitalized[7], and promoted the development of other scientific branches (meteorology, geophysics, astrophysics, nuclear physics, electronics) which were consolidated as institutes during the 1970s (Clark Arxer, 1999; Baracca, Fajer and Rodríguez, 2014, 142-145).

In 1965, following a proposal by Fidel Castro, the National Centre for Scientific Research (CNIC) was created, as a general support structure for the growing research centres and activities, whose main purpose was to promote and support scientific research in all areas, and to develop post-graduate training (Pruna Goodgall, 2006, 285-288; Baracca, Fajer and Rodríguez, 2014, 158-59). Biological sciences would come to form a dominant part of the CNIC, where most members of the highest level scientific staff of today's biotechnological, genetic engineering and pharmaceutical centres have been trained (see further on). It became the generator of some of the most important Cuban research centres (Centre of Neurosciences, National Centre of Animal Health, Centre of Research in Genetics and Biotechnology, Centre of Immunoassay). The CNIC developed collaboration with leading research institutions in the Soviet Union and the socialist countries, and also with the French National Centre for Scientific Research (CNRS), and with Spain and the United States (in the area of neurosciences).

*The basic (unsuspected) support and contributions from "Western" scientists*

---

5 Indeed, Mills' last work was on Cuba (Mills 1960); in it, while denouncing the United States' ignorance of history, including the history of its own imperialism, he stressed that one of the challenges the Revolution would have to face was the lack of well-qualified people.

6 Subtle strategies for the diffusion of these scientific sectors, protecting instead the most sensitive ones for military applications, were implemented, mostly by the United States (Krige 2006; Baracca 2012).

7 Actually, a Royal Academy of Medical, Physical and Natural Sciences of Havana had been created in 1861 (P. M. Pruna Goodgall. 1994. National Science in a Colonial Context: The Royal Academy of Sciences of Havana, 1861–1898. *Isis* 85(3): 412–426). After the establishment of the Republic in 1902, the adjective "Royal" was eliminated. In 1962 for the first time the new Academy acquired a national dimension and an effective role.

As anticipated, the contribution and support from Soviet scientists and institutions to Cuban science could hardly be underestimated. Much less well-known are the contributions of "Western" scientists. In physics their role was complementary, and quantitatively not comparable with the Soviets', but they brought pivotal contributions in some critical moments. In a period of active interest in the Cuban Revolution, crucial initiatives were proposed by the foreign scientists who participated in the Cultural Congress of Havana, in January 1968 (Baracca, Fajer and Rodríguez 2014, 151-154). Advanced Summer Courses were organized in Cuba from 1968 to 1973, in various disciplines. They had important consequences in physics, since the French specialists brought materials and equipment (Cernogora 2014), and gave a decisive turn to Cuban microelectronics, with the change from germanium to planar silicon technology. Actually, the Cuban physicists succeeded in developing the new techniques in a surprisingly short lapse of time, reaching around the mid-1970s an international level in this field, comparable with that of the main much larger Latin American countries which had a longer scientific tradition (Vigil Santos 2014; Baracca, Fajer and Rodríguez 2014, pp.164-168). However, the introduction of high integration microelectronics proved how subalternity can prove hard to overcome, since it cut out all the developing countries from this promising field into which they had ventured. As a consequence, in Cuba several activities had to be reoriented, in particular that concerning optoelectronic sensors.

If in physics the contribution from Western scientists, no matter how important, was after all subsidiary with respect to that of the Soviets, on the contrary in the fields of modern biology the Soviet scientists were seriously behind regarding the modern developments of genetics and molecular biology, and here the role of Western scientists was decisive, although still unreported (Baracca and Franconi 2016, p. 47-50). In fact, a crucial support to the training of Cuban scientists came from Italian specialists. Taking part in the first 1968 Summer School, the director of the group of molecular genetics of the International Laboratory of Genetics and Biophysics of Naples, Paolo ("Pablo") Amati, realized the necessity of providing a sounder base in genetics, and proposed to the Cuban academic authorities more intensive six-month courses. Three courses were in fact held from 1971 to 1973, involving several of the best Italian specialists. These courses trained the new generation of leading Cuban scientists that have had a fundamental role in the establishment of the next Cuban biotechnology industry. In subsequent years some of them participated in further training programmes in Italy, and later in other international research centres, in particular in France, interacting and collaborating with top scientists in the field.

The range of initiatives by Italian specialists in the field of biology was even wider (Baracca and Franconi 2016, p. 47-50). In 1969 an Italian molecular biologist, Bruno Colombo (1936-1989), decided to move from MIT to Cuba, to the Institute of Haematology of the William Soler Hospital, which soon became well known throughout Latin America. He specialized in haematology and haemoglobin, and spent 8 years working in Cuba, where he succeeded in being sent equipment, reagents, documentary research, circumventing the embargo (sometimes even from the US), and developed an instrumental collaboration that led Cuban research in the field at an international level. He also taught in the courses organized by Amati in the section of human genetics. His colleague, the chemist Sandro Gandini, also spent a long time working in Cuba.

Moreover the Italian virologists Giovanni Battista Rossi (1935-1994) and Paola Verani gave summer courses from 1970 for three years, and Sancia Gaetani gave courses on nutrition.

*Stabilization and achievements of a growing advanced scientific system*

Many other institutes and centres were created in Cuba during the 1970s, shaping the basic structure of an advanced scientific system of growing complexity and integration (Baracca, Fajer and Rodríguez 2014, 175-204). The creation of efficient scientific structures, the acquisition of essential equipment and the preparation of a critical mass of trained scientists in the basic scientific fields, made the early successes possible, as regards microelectronics. On the other hand, the growing complexity of the system and the multiplication of activities required new means of central coordination, which replaced the previous initiatives of individual institutions or groups, although the typical Cuban inventiveness has never ceased to play a significant role.

During the 1960s and 1970s, Cuba trained about 1.8 researchers per 1,000 inhabitants, a figure far above the average in Latin America (0.4) and close to that of Europe (2.0) (UN 2001).

> "This situation placed Cuba well outside the trend of correlation between the size of a countries' scientific system compared with that of its economy, and subsequently these favourable conditions permitted a new development programme to be established" (López Mola et al. 2006).

Still, in 1976, at the peak of Cuban-Soviet cooperation, Cuban scientific institutions were also affiliated with forty-eight international scientific organizations, and a further thirty-three applications were being processed (MINSAP 1976: cit. in Reid-Henry 2010, 175).

**The door is opened towards more noteworthy achievements**

During the 1980s more ambitious goals were fixed, and remarkable results accumulated. New research centres were established or grew out of pre-existing groups and collectives, providing them with greater capabilities for applying their scientific results, and existing branches received additional support. International collaboration agreements and participation in joint programmes increased. The Soviet-style planning in the economy adopted by Cuba led to centralizing in the ACC and in the Ministry of Higher Education (MES) respectively the decisions concerning the organization of research activity and higher education, as well as the establishment of new collaboration agreements. Research centres became mainly multidisciplinary. The initial choice of promoting the development of a strong physics sector proved its effectiveness, apart from its brilliant results, since physicists, once exceeding a critical mass, also supplied well-trained scientists to a wide variety of institutions, in which their flexible training allowed them to switch to other related disciplines, where they contributed to developing rigorous methods and approaches.

Actually, Cuban scientists had familiarized with scientific research in a very flexible way. Progress in new fields advanced initially through the (surprisingly quick) learning, adaptation and integration of new technologies to local conditions, but it was soon followed by more in depth research and the swift creation of original developments.

*Challenging projects and results in physics*

The multiple developments and projects in physics in the 1980s cannot be thoroughly examined here (Baracca, Fajer and Rodríguez 2014, 180-204), so just some of the most significant ones will be mentioned.

A challenging programme in the nuclear sector was launched, having as its cornerstone the project to build a nuclear power plant supplied by the Soviet Union in Juraguá (Cienfuegos Province), with the goal of reducing dependence on imported oil (Baracca, Fajer and Rodríguez 2014, 189-193). The introduction of nuclear technologies in the country's economy required, as is always usual, strong centralized organization and control, besides large investments. The Cuban Commission for Atomic Energy (CEAC) was created, responsible for coordinating the main national institutions involved in nuclear research activities, and as advisor to the government. The nuclear programme was developed in close collaboration with the USSR, the European socialist countries and the International Atomic Energy Agency (IAEA). The whole sector was reorganized, and given a special and favoured place and structure. Several institutions were created for the scientific and technological support of the programme. This created some tensions and unbalance inside the Cuban scientific community.

The investment programmes in physics had a great impact also on the advancement of the country's electronics projects (Baracca, Fajer and Rodríguez 2014, 183-189).

The sudden launch in 1987 of a research programme in superconductivity deserves to be mentioned, because it followed a very typical Cuban approach (Arés Muzio and Altshuler 2014). The phenomenon of superconductivity, discovered by Kamerlingh Onnes in 1911 in Leiden, consists in the electric resistance of certain materials vanishing to zero below a characteristic critical temperature of a few degrees above absolute zero (-273 ºC), precisely below the liquefaction temperature of helium (-269 ºC). No experimental research in this field had been previously carried out in Cuba, since liquid helium was not available in the country, but only liquid nitrogen (-195,80 °C). But in 1986 G. Bednorz and A. Müller discovered a compound with a higher critical temperature (-248 ºC, however still below the liquefaction temperature of nitrogen). For this discovery, they were awarded the Nobel Prize. Less than 6 months later, Chu and his collaborators reached in a specific ceramic material a critical temperature above -183 °C, a temperature higher than the liquefaction temperature of nitrogen. Laboratories worldwide were now able to reproduce this finding. In Cuba, at the magnetism laboratory in Havana, Oscar Arés Muzio, that had accumulated long experience with ceramic materials, succeeded with his collaborators in creating a high temperature superconductor barely two months after Chu's announcement. Such a world-standard scientific result, obtained from no previous experience in the specific field, received widespread coverage in the national media and Fidel Castro's direct interest. The group was endowed with resources for setting up a new superconductivity laboratory, which attracted many brilliant young researchers and developed high standard research.

*The appearance and surprising growth of an advanced biotechnology sector*

Most visible among the transformations of the 1980s was the emergence in Cuba of a high-tech production sector associated with biotechnology, the medical-pharmaceutical industry, and the health care system. The absolute originality of this process was that, on the one hand it grew in an independent way from the Soviet Union, but on the other hand it neither followed the model of organization nor the approach of the capital-intensive multinational big biotechnological industry (Baracca and Franconi, 2016). Indeed,

> "…biotechnology in Cuba was driven by public health demand. … In place of an economic imperative, therefore, were the social demands that the Cuban scientists apply to the product of their research as soon as possible. … their model of fast science was … very much against the global trend of the time" (Reid-Henry 2010, 18-19).

This outstanding and original outcome may even be seen, almost paradoxically, as a side effect of US policy, since Cuba´s dependence on pharmaceuticals and health supplies from other countries drove Castro´s decision to launch, when the field of biotechnology was still in its infancy in the world, a series of programmes aimed at applying these new techniques to the health care sector in order to address domestic health problems. This choice could obviously avail itself of an efficient and widespread health system, medical schools, research institutes, and factories of pharmaceutical and medical products.

In that period, some larger Latin American countries, with an older tradition, also set up programmes for the development of biotechnology (Peritore and Galve-Peritore 1995), but it was the small island that achieved the most amazing results, and in a surprisingly short time. Cuba had an adequate number of trained biologists from the early intensive courses of the 1970s, who subsequently specialized in Italy, France and other countries.

In 1981 the *Frente Biológico* (Biological Front) was created, in order to coordinate biology centres, groups and specialists, and to ensure a connection between scientific development and the national economy, under the supervision of the Cuban Council of State. The government's long-term vision in supporting a sector not likely to produce short-term returns for the investments highlights a constant feature of Cuban biotechnology, and in fact it turned out to be a key to Cuban success in the field of health biotechnology (Thorsteinsdóttir et al. 2004).

The initial plan consisted of the rapid acquisition of international know-how and technologies and their translation into useful products. Around 1980 a group of Cuban physicians took an interest in the recent application of interferon for treating cancer, an illness that took priority as a public health issue, besides heart diseases, after endemic illnesses had been eradicated in Cuba (Cooper et al.

2006; Health in the Americas 2012, 244)[8]. These physicians first visited one of the most eminent American oncologists, Randolph Lee Clark (1906-1994), director of the MD Anderson Cancer Hospital in Houston, who in turn visited Cuban health facilities and met Fidel Castro, leaving him convinced that interferon was the right drug to develop (Feinsilver 1993a; Reid-Henry 2010, 13-19). Clark directed the Cubans to Kari Cantell in Helsinki, who had developed a technique for making useful amounts of industry-standard interferon from human blood cells.

As Cantell himself relates in detail (Cantell 1998, 141-153), at the beginning of 1981 a group of six Cubans, virologists, immunologists and biochemists visited Cantell's laboratory for one week in order to learn the technique. Once back in Cuba, with Castro's direct interest, a mansion was converted into a laboratory, and in barely a couple of months the Cuban scientists reproduced the process and stabilized the production of interferon. For the first time Cuba entered an industrial sector at the very moment it was being born globally. Moreover, in the same year 1981 Cuban doctors used interferon in medical practice during a virulent epidemic of haemorrhagic dengue fever showing again an open-minded and problem-solving approach as well as representing an early example of the so called pharmaceutical 'repurposing'. The close link between research and clinical testing and application would remain a peculiar feature of the Cuban biomedical system.

*The development of an advanced biotechnology industry, to increase sovereignty*

The success in producing and using interferon posed the urgent need to produce greater quantities, and led to the decision of creating the Centre for Biological Investigation (CIB) which was built in only six months, and was to become a part of the Western Scientific Pole of Havana, a huge scientific complex consisting at present of 52 major integrated research, education, health, and economic institutions devoted to the biotechnology field. The initial work on the purification of interferon was complemented by a parallel project to produce it by genetic engineering, a result that few others had obtained. The orientation towards genetic engineering was not driven in Cuba by the logic dominant in Western industry, or by the search for cutting edge scientific results, but by the fact that it did the job, responding to national needs. Molecular biologists were immediately sent to the Pasteur Institute in Paris. In 1984 they succeeded in developing a whole new approach from Cantell´s technique, a second-generation, recombinant interferon (Reid-Henry 2010, 46-47). Between 1982 and 1986 the development of molecular biology and genetic engineering at the CIB represented the first step leading to their own innovations and development of knowledge.

> "By 1986 Cuba was 'the second-largest producer of natural human leukocyte interferon, after Finland' … Interferon was chosen as a model to develop genetic engineering and biotechnology techniques ….and, …. it served as a model for the development of advanced molecular biology skills … A U.S. biotechnology industry analyst substantiated the Cuban approach when he suggested in 1990 that 'alpha interferon almost serves as a paradigm for all of these biological response modifiers …' which are at the forefront of biotechnology research" (Feinsilver 1995, 101).

> "The quality of Cuban equipment and research facility ... was of Japanese or West European level, but due to the U.S. trade embargo it was very costly ... Consequently, Cuban scientists had to learn to produce their own restriction enzymes, make tissue cultures, establish virus collections, as well as to develop and manufacture equipment to do electrophoresis and gas chromatography" (Feinsilver 1995, 103).

In 1986 the Centre for Genetic Engineering and Biotechnology (CIGB) was inaugurated, followed by the Finlay Institute (opened in 1987) and the Centre for Molecular Immunology (CIM, opened in 1991). In the following years the range of interests and activities of Cuban biotechnologists considerably widened, in connection with the public health system and many protein-based pharmaceuticals were produced, including cytokines, monoclonal antibodies and vaccines. Interestingly, about the latters, all children in Cuba are vaccinated against hepatitis B and meningococcal meningitis (BC), two modern recombinant vaccines, both developed in Cuba; the latter is the only internationally marketed vaccine available for serogroup B meningococcus and won the World Intellectual Property Organization's gold medal for innovation (Reid-Henry 2010, 64-66). During the years 1989-90, over 96% of Cuban population groups at risk were vaccinated, with the usual close correlation between research and clinical testing.[9] Orders for this vaccine immediately arrived from Brazil, Argentina, Colombia and other countries.

---

[8] Vaccination in Cuba, begun in 1962 under the National Immunization Program (NIP,) considerably reduced the infectious disease burden contributing to elimination of diseases such as: poliomyelitis (1962); diphtheria (1979); measles (1993); pertussis (1994); and rubella (1995) (Reed et al. 2007). It is worth noting that the current vaccine schedule targets all Cuban children for immunization against 13 diseases with 11 vaccines, eight of which are produced by the country's Scientific Pole.

> "Cuban research also prioritizes developing affordable vaccines for diseases affecting poor populations, such as typhoid fever and cholera: a fundamentally needs-driven, rather than market-driven approach. This can be contrasted with transnational pharmaceutical companies, which have come under increasing criticism for placing market interests before global health solutions, resulting in investment of 90% of R&D dollars worldwide in developing treatments for diseases affecting the 10% of the world's population that can afford the results" (Evenson 2007).

The 'full-cycle' conception was an explicit strategy, made easier by centralized state control, as many commentators have remarked (Elderhost 1994; Thorsteinsdóttir et al. 2004; Reid-Henry 2010). The development of a national capacity in biotechnology was seen as a strategy to increase sovereignty and independence from the transnational companies of the industrialized countries, especially in the medical sector. In the 1980s Cuba was already acting in the major markets in Eastern Europe and the former Soviet Union, attempting to promote technology transfer within the COMECON, an alliance of countries that did not recognize Western intellectual property laws, but it also tried to increase scientific relations with the West (Reid-Henry 2010, 55-56; de la Fuente 2001), a strategy that proved to be useful after the unforeseen collapse of the Soviet Union.

*A country of men and women of science, lifting themselves out of subalternity*

Summing up, during the 1980s Cuba reached the aim of overcoming the condition of subalternity becoming indeed "a country of men and women of science", a reservoir of intellectual resources and know-how both for socialist and developing countries, upholding its freedom of action in the international arena. The Cuban scientific system reached its "adult" stage, grew considerably stronger, larger, and more articulated, and acquired a substantial autonomy in its organization and development. In particular, the biomedical complex became one of the main sources of hard currency for the country. In the field of science autonomy is not autarky, but the capability of interacting, exchanging and collaborating on an equal footing with the more developed, Eastern and Western, countries (including even, in some relevant cases, the United States).

**Cuba faces a new challenge**

However, the overcoming of the condition of subalternity and the attainment of full autonomy, even in a specific sector, is never a final result, guaranteed once and forever: not even in the case of big and powerful countries, as was demonstrated by the sudden and unexpected collapse of the Soviet Union. Much less in the case of Cuba, for which the consequences were truly dramatic, and put at risk the conquest of three decades, if not the subsistence of the regime, as many analysts foresaw. But Cuba is still there, and keeps representing a benchmark for the fresh breeze that in the past two decades is blowing over Latin America and the Caribbean.

How did Cuba succeed in this second challenge?

*A "disaster proof" scientific system*

In fact, in that terrible circumstance the Cuban scientific system demonstrated the solidity it had reached, substantially resisting the shock. Of course, the consequences were serious, scientific activities were profoundly affected. In physics and in other scientific sectors several activities declined, or even had to be closed down or redirected (Baracca, Fajer and Rodríguez 2014, 203-208;

---

9 Reid-Henry (p. 66-67) discusses an interesting comparison of the practices of research and clinical testing in the case of this vaccine adopted by Cuba and by Norway.

Baracca and Franconi 2016, Chapter 5). For instance, the project of the nuclear power plant in Juraguá was abandoned after the failure to find alternative international partners. Experimental activities especially encountered great difficulties due to the shortage of spare parts for equipment of Soviet fabrication. For its part, the United States tightened the economic blockade with even more restrictive laws (Torricelli Act, 1992, and Helms-Burton Act, 1996)[10]. Such a stranglehold on Cuba's economy worsened shortages of food and medicine (Kirkpatrick 1996). Irreplaceable scientific information and updating had to be cut, due to the prohibitive subscription fees of international journals.

But the Cuban scientific system did not collapse, demonstrating the toughness, cohesion and level of self-sufficiency it had reached. In this critical circumstance the ability of Cuban scientists (not to say of Cubans in general) to make a virtue of necessity and to exploit their own resourcefulness proved once more successful.

The widespread international collaboration established by the Cuban scientific community throughout the previous decades proved all the more useful in the new situation, since it provided alternative options and invaluable opportunities (see Baracca, Renn and Wendt 2014, Part Three; Baracca and Franconi 2016, Chapter 5). The physicists intensified their relations and exchanges with Spain, Mexico, Brazil and other countries, besides those with the International Centre of Theoretical Physics in Trieste.

*Support to advanced science proves to be once again the trump card*

Once again the renewed challenge was met by relying on science as a driving force of economic recovery and reaffirmation of the autonomy and independence of the country. In order to keep the Cuban economy and the Revolution afloat Castro chose to prioritize three economic sectors for investments: tourism, food production, and biotechnology and medical exports.

> "Faced with economic calamity, Castro did something remarkable: he poured hundreds of millions of dollars into pharmaceuticals" (Starr 2012).

Although the difficult economic situation made it impossible to provide equal, or even comparable, support to all scientific branches.

Obviously the situation was radically different from three decades before, both domestically and internationally. Even the most favourable analysts wondered if this choice would prove successful: "Can biotechnology save the revolution?" (Feinsilver 1993b). Could Cuba find sufficient markets and ways to commercialize its products in the globalized capitalistic market, controlled by transnational pharmaceutical companies with patents, the control of markets, and enormous capital and finances? Intellectual property (IP) rights were one more obstacle. The Cuban economy had adapted to guaranteed and protected markets for standardized products, lacking the level of efficiency and competitiveness of capitalist markets, with their specific commodity-related rules of marketable quality.

Once more, and in very critical conditions, Cuba found itself on the border of the continually growing technical gap between the poor and the rich countries, between subalternity and hegemony. And once more the Cuban government's courageous choice seems to have been a trump card (Baracca and Franconi 2016, Chapter 5).

Activities in the biomedical field were confirmed and even reinforced. From 1990 to 1996, the Cuban government invested around 1 billion US dollars in the Western Havana Bio-Cluster, the foremost Cuban biomedical complex.

> "According to Elena Siméon, the director of the *Academía de Ciencias de Cuba* (ACC), approximately 8,000 people worked in scientific research relating to biotechnology in 200 research institutes in 1993. In the period 1988-1992, more than US $300 million was invested in medical and pharmaceutical industrial biotechnology" (Elderhost 1994).

---

10 The deeply controversial Helms-Burton Act extended the prohibition of trade with Cuba to companies doing business with it and to companies that use property Cuba had nationalized in 1960 from American companies. The US Government prosecuted Merck, the largest pharmaceutical firm in the USA, for an exchange of scientific information with Cuba.

*Addressing new challenges*

With open-mindedness and creativity Cuba addressed the complex problems of a completely new nature posed by the forced shift from the socialist to the capitalistic market. Besides the regulation and standardization of products, the main obstacle was represented by intellectual property (IP) rights. Cuba's previous form of socialist regulation of IP was the expression of public rather than private interests (intellectual property to the author, vs. commercial property to the state). In 1986 Cuba had strongly opposed the new global IP rights regime in the Uruguay Round of General Agreement on Tariffs and Trade (GATT). The developing countries resisted accepting the TRIPS agreement. Now Cuba had no choice but to integrate into the capitalistic system of IP, and in 1995 signed on to TRIPS (Sanchelima 2002; Càrdenas 2009; Plahte and Reid-Henry 2013). But once again it was true to itself, integrating into the international system while keeping margins of typical Cuban flexibility, resorting to loopholes in the embargo, and negotiating bilateral agreements with specific countries when possible. In particular, it found an original solution to IP in state ownership.

> "The patents of the Cuban industry are owned by the government agency, which avoids the problem of mutual blocking. This agency functions as a kind of patent pool, where every firm has the possibility of using complementary knowledge in advancing new products. … This resembles more an *internal* open source of innovation [coherent with] the notion of cooperation instead of competition" (Càrdenas 2009).

*Cuba beyond the obstacles*

Almost a decade later, one study concludes that

> "Ten of these institutions [of the Western Havana Bio-Cluster] are at the core of the system as they supply economic support to the whole effort with their production capacities and exports. They are performing more than 100 research projects which have generated a product pipeline of more than 60 new products most of which are protected by intellectual property, and more than 500 patents have been filed overseas" (López Mola et al. 2006).

The Economist Intelligence Unit estimated that the increase in non-tourism services exports between 2003 and 2005 was around US $1.2 billion for a total of US $2.4 billion, which put non-tourism services ahead of gross tourism earnings (of US$2.3 billion) in 2005. Most of these were medical services (Feinsilver 2006). And a report of the Word Bank states,

> "[...] the growth of the local pharmaceutical industry, which by the mid-1990s was bringing Cuba some 100 million dollars a year in export earnings, has not only covered domestic demand for medicines, but has also led to the development of products that compete on the international market. Cuba is the only country in the world, for example, that has come up with an effective vaccine against meningitis B" (Kaplan and Laing 2005).

The differentiation of Cuban biomedical products for the international market grew:

> "the meningitis B vaccine ... was the first of its kind in the world ... A more recent example of success is the world's first human vaccine with a synthetic antigen for *Haemophilus influenzae* type b (Hib). ... Ongoing work includes research on recombinant Dengue vaccine, preventative and therapeutic AIDS vaccines, cholera vaccine and a cancer therapeutic vaccine[11]. The sector has also successfully produced diagnostic tests and therapeutics ... In addition, Cuba is developing natural products based on the island's flora. …researchers in Cuba have filed about 500 patent applications in the health biotechnology sector based on more than 200 inventions filed in several countries throughout the world, including the United States, Europe, Brazil, India, China and South Korea. Cuba exports biotechnology products to more than 50 countries" (Thorsteinsdóttir et al. 2004).

In the end,

---

[11] An update of novel therapeutics and vaccines that are currently being evaluated in Cuba in clinical trials or that are in the market was done at the 'First International Convention IMMUNOPHARMACOLOGY-VACCIPHARMA 2015 (III Int. Congress on Immunopharmacology + III International Congress on Pharmacology of Vaccines), Varadero, Cuba, June 14-19, 2015. http://www.immunovaccipharmacuba.com

> "in the wake of the Soviet collapse, Cuba got so good at making knock-off drugs that a thriving industry took hold. Today the country is the largest medicine exporter in Latin America and has more than 50 nations on its client list. Cuban meds cost far less than their first-world counterparts, and Fidel Castro's government has helped China, Malaysia, India, and Iran set up their own factories: 'south-to-south technology transfer'" (Starr, 2012).

*A peculiar industrial structure, "an aberration in today's world"*

In evaluating these outcomes of the Cuban biomedical sector in the highly concentrated and competitive globalized market, one must always take into account the peculiar structure of Cuban industry (Baracca and Franconi, 2016, Chapter 5).

> "Dr Rolando Pérez, one of the top scientists at CIM-Centre of Immunology, points out that the Cuban biotechnology model is 'completely different' from the development of Western pharmaceutical products. 'Pure scientific research, innovation and product development, production and marketing are all integrated under the same roof, or at least in the same institution.' … In a country where there are no private hospitals, and all pharmaceuticals are publicly owned, inevitably all investment comes from the state. … Orthodox Western economists would tend to dismiss this socialist model of medical innovation and production as a quaint aberration in today's world, clearly out of synch in the globalized economy. But the Cuban record boasts 26 inventions with more than 100 international patents already granted" (Fawthrop 2004).

Concerning the Cuban scientific system on the whole, such a targeted effort in the biomedical sector in that difficult economic situation could do nothing other than make imbalances between different scientific sectors grow.

> "During the Special Period, the universities suffered more cutbacks than the research institutions in the health biotechnology sector, and some university centres struggled to do research. Research collaboration between the universities and the research institutions has been encouraged, and some university institutions have made impressive contributions to health biotechnology. For example, researchers from the Faculty of Chemistry at the University of Havana made a leading contribution in the development of the synthetic *H. influenzae* type b vaccine. Universities and research institutions also collaborate in the research training of students" (Thorsteinsdóttir et al. 2004).

Even a rather critical report that describes the crude reality of the situation in 2003, insisting in particular on existing contradictions (Giles 2003), had to recognize that:

> "Given the hardships suffered by researchers outside the charmed circle of priority applied research projects, it is surprising not to hear more complaints from Cuban researchers. Government control may be one factor; open dissent is a risky policy in a non-democratic country. But equally important is an awareness that Cuba has battled against the odds to avoid the chaos and privations suffered by neighbouring countries such as Haiti. Older Cuban scientists, who remember the right-wing dictatorship that preceded Castro, are especially proud of what´s been achieved" (Giles 2003).

On the other hand, the burden constituted in this difficult situation by the anachronistic US embargo could hardly be over-emphasized.

> "The US trade embargo has limited the economic options for Cuba, including development of the health biotechnology sector. For example, Cuba is forced to import research equipment from countries other than the United States—a situation that not only consumes time but adds to the cost. Another challenge imposed by the poor Cuba-US relations is the increasing difficulty that Cuban scientists face obtaining visas to enter the United States to attend conferences and other related activities. Also, …..the uncertainties of the embargo have made it difficult for Cuban papers to be accepted in US journals. The embargo therefore restricts the knowledge flow involving Cuban scientists in the international scientific community and adds costs, because Cubans have to attend conferences that are held in countries other than the United States. Another challenge is the dominance of US firms in the global health biotechnology sector. This may limit the options for Cuba in developing joint ventures, strategic alliances and licensing of their technologies" (Thorsteinsdóttir et al. 2004).

Until a recent past, Cuba's success proved troublesome for the United States, as it emerged from some controversial reciprocal allegations, of a political nature, of biological warfare. In 2002 John R. Bolton, under-Secretary of State for arms control in the Bush Jr. Administration, caused a sensation accusing Cuba of producing small quantities of germs that could be used in biological warfare.[12] The CIA itself later rebuffed these claims.[13]

**Conclusion**

The striking Cuban achievements in science defy the relentless embargo, and even stun and humiliate the nearby powerful and proud enemy. No less politically unwelcome is the well-known Cuban international engagement and "medical diplomacy". Cuba does not miss in fact a single opportunity to offer and supply medical support and disaster relief assistance, including

> "an offer to send over 1000 doctors as well as medical supplies to the United States in the immediate aftermath of Hurricane Katrina. Although the Bush administration chose not to accept the offer, the symbolism of this offer of help by a small, developing country that has suffered forty-five years of US hostilities, including an economic embargo, is quite important" (Feinsilver 2006).

As the two foreign secretaries of the Cuban and the US Academies of Science have jointly remarked in *Science*,

> "Cuba is scientifically proficient in disaster management and mitigation, vaccine production, and epidemiology. Cuban scientists could benefit from access to research facilities that are beyond the capabilities of any developing country, and the U.S. scientific community could benefit from high-quality science being done in Cuba. For example, Cuba typically sits in the path of hurricanes bound for the U.S. mainland that create great destruction, as was the case with Hurricane Katrina and again last month with Hurricane Ike. Cuban scientists and engineers have learned how to protect threatened populations and minimize damage. (Jorge-Pastrana and Clegg, 2008).

Recently, in May 2015, a team of medical specialists was sent to Nepal after the fatal earthquake while, last year, Cuba sent 256 doctors to West Africa (Liberia) where the Ebola outbreak was the most severe. The doctors returned to Cuba in March 2015, after spending six months treating patients affected by the deadly virus. *The New York Times* reported:

> "Cuba stands to play the most robust role among the nations seeking to contain the virus [ebola].
> … only Cuba and a few nongovernmental organizations are offering what is most needed: medical professionals in the field."[14]

The recent re-establishment of diplomatic relations between US and Cuba (initiated at the end of 2014 followed by re-opening of embassies in both countries in July 2015) together with the removal of Cuba from the terrorist list (April 2015) and removal of travel restrictions, now gives the hope that more than 5 decades of policy hostility and subversion will be replaced by a new a policy of engagement and cooperation. While the embargo is still there, President Obama's visit to Havana in March 2016 will remain one of the historical dates of the 21st century. Hopefully, all these recent events, in the near future, will open doors for reciprocal (and global) scientific and medical advances and broaden the acknowledgment of Cuban innovation in healthcare, including the lesson of 'how to do more with less'.

**Acknowledgments**

AB is indebted to the Director of the Max Planck Institute for the History of Science in Berlin, Prof. Jürgen Renn, for his interest, hospitality and support during the beginning of this research.

---

12 J. Miller, Washington Accuses Cuba of Germ-Warfare Research, *The New York Times*, May 7, 2002, http://www.nytimes.com/2002/05/07/international/americas/07WEAP.html. Fidel Castro, CUBA: 'Our weapons are morality, reason and ideas', May 22, 2002, https://www.greenleft.org.au/node/27449. Thinktank disputes Bush administration claims of biowar development in Cuba, Center for International Policy, May 2002, http://www.afn.org/~iguana/archives/2002_05/20020508.html.
13 Wayne S. Smith , More Empty Charges, April 7, 2004, http://articles.sun-sentinel.com/2004-04-07/news/0404060324_1_cuba-biological-weapons-bolton-s-statement. L. and S. San Martin, CIA rebuffs John Bolton and Otto Reich claim of Cuba's biological warfare capabilities, *Miami Herald*, April 09, 2005, http://havanajournal.com/politics/entry/cia_rebuffs_john_bolton_and_otto_reich_claim_of_cuba_biological_warfare_cap/. For a full account see: NTI, Country Profiles, Cuba, Biological, http://www.nti.org/country-profiles/cuba/biological/.
14 The Editorial Board, "Cuba's impressive role on Ebola", *The New York Times*, 19 October 2014, http://nyti.ms/1t0LpUn.

All the information on the development of physics in Cuba originates, and is quoted, from previous comprehensive works which were published in the volumes: A. Baracca, J. Renn and H. Wendt (eds.), *The History of Physics in Cuba*, Berlin, Springer, 2014, and A. Baracca and Rosella Franconi, *Subalternity vs. Hegemony, Cuba's Outstanding Achievements in Science and Biotechnology,* Berlin, Springer, 2016.

We are indebted with Carlos Cabal that suggested the research as well as with Paolo Amati for his witness and help that stimulated further research, still in progress, on the role of Italian geneticists on the development of Cuban biotechnology. We are grateful to Marina and Luciano Terrenato, University of Rome "La Sapienza" for proving us information and original documents about the 1971 'Summer school' in Genetics, and to the haematologist Gisela Martinez for information about Bruno Colombo.

RF is grateful to Cuban colleagues, in particular the scientists of the International Centre of Genetic Engineering and Biotechnology (CIGB), who have evoked her admiration and represent an example of being public-sector researchers.

**References**

Altshuler José, and Angelo Baracca. 2014. The teaching of physics in Cuba from colonial times to 1959. In *The History of Physics in Cuba*, eds. Angelo Baracca, Jürgen Renn and Helge Wendt, 57-106. Berlin: Springer.

Arés Muzio Oscar, and Ernesto Altshuler. 2014. Superconductivity in Cuba: Reaching the frontline. In *The History of Physics in Cuba*, eds. Angelo Baracca, Jürgen Renn and Helge Wendt, 301-306. Berlin: Springer.

Arias De Fuente Olimpia. 2014. An interview with professor Melquíades De Dios Leyva, December 2008. In *The History of Physics in Cuba*, eds. Angelo Baracca, Jürgen Renn and Helge Wendt, 285-288. Berlin: Springer.

Baracca Angelo, Victor Fajer Avila, and Carlos Rodríguez Castellanos. 2014. A comprehensive study of the development of Physics in Cuba from 1959. In *The History of Physics in Cuba*, eds. Angelo Baracca, Jürgen Renn and Helge Wendt, 115-234. Berlin: Springer.

Baracca Angelo, Jürgen Renn, and Helge Wendt (Eds.). 2014. *The History of Physics in Cuba*. Berlin: Springer.

Baracca Angelo, and Rosella Franconi. 2016. *Subalternity vs. Hegemony, Cuba's Outstanding Achievements in Science and Biotechnology*, *1959-2014.* Series Title: SpringerBriefs History Science Technology. ISBN: 978-3-319-40608-4, http://www.springer.com/us/book/9783319406084

Buckley Jonathan, Jorge Gatica, Mark Tang, Halla Thorsteinsdóttir, Alok Gupta, Sabine Louët, Min-Chol Shin and Mark Wilso. 2006. Off the Beaten Path. *Nature Biotechnology* 24: 309-315.

Cantell Kari. 1998. *The Story of Interferon: the Ups and Downs in the Life of a Scientist*. World Scientific.
Cárdenas Andrés. 2009. The Cuban Biotechnology Industry: Innovation and universal health care. www.theairnet.org/files/research/cardenas/andres-crdenas_cubab_biotech_paper_2009.pdf.

Castro Fidel. 1960. *El Futuro de nuestra Patria tiene que ser necesariamente un Futuro de Hombres de Ciencia*. National Agrarian Reform Institute (INRA), Havana.

Cernogora Jaqueline. 2014. A witness to French-Cuban cooperation in physics in the 1970s. In *The History of Physics in Cuba*, eds. Angelo Baracca, Jürgen Renn and Helge Wendt, p. 365-79. Berlin: Springer.

Clark Arxer Ismael 1999. *138 Años de la Academia de Ciencias de Cuba: Visión de la Ciencia y del Proceso Histórico Cubano*. Editorial Academia. Havana.

Clark Arxer Ismael. 2010. Cuba. In *UNESCO Science Report 2010*, Chapter 6, 123–1331. http://www.unesco.org/new/en/natural-sciences/science-technology/prospective-studies/unesco-science-report/unesco-science-report-2010.

Cooper Richard S., Joan F. Kennelly and Pedro Orduñez-García. 2006. Health in Cuba. *International Journal of Epidemiology*. 35 (4): 817-824.

de la Fuente José. 2001. Wine into vinegar: the fall of Cuba's biotechnology. *Nature Biotechnology*. 19: 905-7.

Editorial. 2009. Cuba's biotech boom. The United States would do well to end restrictions on collaborations with the island nation's scientists. *Nature*. 457 (January): 8

Elderhost Miriam. 1994. Will Cuba´s biotechnology capacity survive the socio-economic crisis? *Biotechnology and Development Monitor*. 20 (September). 11-13/22.

Evenson Debra. 2007. Cuba´s Biotechnology Revolution. *MEDICC Review*. 9 (1): 8-10.

Fawthrop Tom. 2004. Cuba Ailing? Not Its Biomedical Industry. The Straits Times, 26 January 2004. http://yaleglobal.yale.edu/content/cuba-ailing-not-its-biomedical-industry

Feinsilver Julie M. 1993a. Healing the Masses. Cuban Health Politics at Home and Abroad. Berkeley, CA: University of California Press.

Feinsilver Julie M. 1993b. Can biotechnology save the revolution? *NACLA Report of the Americas*. 21 (5):7-10.

Feinsilver Julie M.1995. Cuban Biotechnology: the Strategic Success and Commercial Limits of a First World Approach to Development. In *Biotechnology in Latin America: Politics, Impacts and Risks*. Peritore N. Patrick and Galve-Peritore A. Karina (Eds). Wilmington, D.E.: Scholarly Resources, 1995.

Feinsilver Julie M. 2006. La Diplomacia Medica Cubana: Cuando La Izquierda Lo Ha Hecho Bien. Foreign Affairs. 6 (4):81-94 (English transl: Cuban Medical Diplomacy: When the Left Has Got It Right. http://www.coha.org/cuban-medical-diplomacy-when-the-left-has-got-it-right/.

Fink Gerald R, Alan I. Leshner, and Turekian Vaughan C. 2014. Science diplomacy with Cuba. *Science*. 344 (6188). 1065.

Giles Jim. 2005. Cuban Science: ¿Vive la revolution?. *Nature*, 436, 21 July 2005, 322–324.

Health in the Americas. 2012 Edition. Country Volume *Cuba*. Pan American Health Organization.

Hoffmann Bert. 2004. *The Politics of the Internet in Third World Development. Challenges in Contrasting Regimes with Case Studies of Costa Rica and Cuba.* New York, Routledge.

Huberman Leo, and Paul M. Sweezy. 1960. *Cuba: Anatomy of a revolution*. New York: Monthly Review Press.

Jorge-Pastrana Sergio, and Michel T. Clegg. 2008. US-Cuban scientific relations. *Science*, 322, 17 October, p. 345.

Kaiser Jocelyn. 1998. Cuba's billion-dollar biotech gamble. *Science*. 282(5394): 1626–1628.

Kaplan Warren, and Richard Laing. 2005. Local Production of Pharmaceuticals: Industry Policy and Access to Medicines. Health, Nutrition and Population Discussion Paper. The World Bank. Jan 16.

Kellner Douglas. 1989. *Ernesto "Che" Guevara (World Leaders Past & Present)*. Chelsea House Publishers. Library Binding edition.

Kiirkpatrick Anthony F. 1996. Role of the USA in the shortage of food and medicine in Cuba. *The Lancet*. 348: 1489-91.

Krige John. 2006. *American Hegemony and the Postwar Reconstruction of Science in Europe*. Cambridge MA: MIT Press.

López Mola Ernesto, Ricardo Silva, Boris Acevedo, José A. Buxadó, Angel Aguilera, and Luis Herrera. 2006. Biotechnology in Cuba: 20 years of scientific, social and economic progress. *Journal of Commercial Biotechnology*. 13: 1-11.

López Mola Ernesto, Ricardo Silva, Boris Acevedo, José A. Buxadó, Angel Aguilera, and Luis Herrera. 2007. Taking stock of Cuban biotech. *Nature Biotechnology*. 25 (11, November): 1215-6.

Mills Charles. 1960. *Listen, Yankee: The revolution in Cuba*. New York: Ballantine Books.

MINSAP. 1976. *Informe Anual*. Havana.

Pérez Rojas, Hugo 2014. Interview by A. Baracca: The rise and development of physics in Cuba: an interview with Hugo Pérez Rojas in May 2009. In Baracca, Renn, and Wendt, 2014. p. 279-284.

Peritore N. Patrick, and Galve-Peritore A. Karina (Eds). 1995. *Biotechnology in Latin America: Politics, Impacts and Risks*. Wilmington, D.E.: Scholarly Resources.

Plahte Jens and Simon Reid-Henry. 2013. Immunity to TRIPS? Vaccine Production and the Biotechnology Industry in Cuba. In: Löfgren H. and Williams O. D. (Eds). 2013. *The New Political Economy of Pharmaceuticals: Production, Innovation and TRIPS*. Pelgrave Macmillan. 70-90.

Pruna Goodgall Pedro M. 1994. National Science in a Colonial Context: The Royal Academy of Sciences of Havana, 1861–1898. *Isis* 85(3): 412–426.

Reid-Henry Simon. 2010. *The Cuban Cure: Reason and Resistance in Global Science.* Chicago: University of Chicago Press.

Sáenz Tirso W., and García-Capote Emilio. 1989. *Ciencia y Tecnología en Cuba*. Editorial in Ciencias Sociales. Havana.

Sanchelima Jesús. 2002. Selected aspects of Cuba´s intellectual property laws. Cuba in Transition: Volume 13. Association for the Study of the Cuban Economy (ASCE) 213-219. http://www.ascecuba.org/publications/proceedings/volume12/pdfs/sanchelima.pdf.

Sartre Jean-Paul. 1961. Sartre on Cuba. New York: Ballantine Books.

Starr Douglas 2012. The Cuban Biotech Revolution. http://www.wired.com/wired/archive/12.12/cuba_pr.html.

Thorsteinsdóttir Halla, Tirso W Sáenz, Uyen Quach, Abdallah S Daar and Peter A Singer. 2004. Cuba. Innovation through synergy. *Nature Biotechnology*. 22 (Supplement) December: 19-24.

U.N. 2001. *Human development report 2001: Making new technologies work for human development. United Nations Development Programme*. Oxford University Press, Oxford, UK.

Vigil Santos Elena. 2014. Experimental semiconductor physics: the will to contribute to the country's economic development. In *The History of Physics in Cuba,* eds. Angelo Baracca, Jürgen Renn and Helge Wendt, 289-294. Berlin: Springer.